\documentclass[
tightenlines,
aps,
twocolumn,
nofootinbib,
floatfix]{revtex4}

\usepackage{graphicx}
\usepackage{float}

\newcommand{\be}{\begin{equation}}
\newcommand{\ee}{\end{equation}}
\newcommand{\ba}{\begin{eqnarray}}
\newcommand{\ea}{\end{eqnarray}}

\newcommand{\slsh}{\rlap{$\;\!\!\not$}}     

\newcommand{\gsim}{\mbox{\raisebox{-0.3ex}{%
\footnotesize $\:\stackrel{>}{\sim}\:$}} }

\begin{document}
\begin{titlepage}

\begin{flushright}
\vbox{
\begin{tabular}{l}
\end{tabular}
}
\end{flushright}

\vspace{0.6cm}

\title{$W+3$ jet production at the LHC as a signal or background}

\author{Kirill Melnikov \thanks{
e-mail:  kirill@phys.hawaii.edu}}
\affiliation{Department of Physics and Astronomy,\\
Johns Hopkins University,
Baltimore, MD, USA}
\author{Giulia Zanderighi  \thanks{
e-mail: g.zanderighi1@physics.ox.ac.uk}}
\affiliation{Rudolf
Peierls Centre for Theoretical Physics, \\
1 Keble Road, University of
        Oxford, UK}

\begin{abstract}
\vspace{2mm}

We discuss the production of $W$ bosons in association with three jets
at the LHC. We investigate how next-to-leading order QCD 
corrections modify basic kinematic distributions of jets and
leptons. We also address the magnitude of NLO QCD effects in $W+3~{\rm
  jet}$ observables, relevant for SUSY searches at the LHC. 

\end{abstract}

\maketitle

\thispagestyle{empty}
\end{titlepage}

\section{Introduction}

A good understanding of complicated multi-particle processes is
important for the LHC physics.  To achieve this goal it is useful to
have next-to-leading order (NLO) QCD predictions for such processes
(see e.g. Ref.~\cite{Bern:2008ef}).  In the past, {\it three} final
state particles was the highest multiplicity for which NLO QCD
computations were feasible, but this changed this year when four
groups~\cite{Bredenstein:2008zb,Bredenstein:2009aj,Ellis:2009zw,Ellis:2009bu,Berger:2009ep,Berger:2009zg,Bevilacqua:2009zn}
reported first results on NLO QCD corrections to processes with {\it
four} particles in the final state. To arrive at these results a
variety of methods including highly-refined Passarino-Veltman
reduction algorithm~\cite{Passarino:1978jh,Denner:2005nn},
Ossola-Pittau-Papadopoulos (OPP) method
\cite{Ossola:2006us,Ossola:2007ax} and unitarity
techniques~\cite{Bern:1994zx,Bern:1994zxy,Bern:1998zxyw,Britto:2004nc,Ellis:2007br,Giele:2008ve}
were used.  Successful completion of these computations and a 
large number of one-loop amplitudes with six and more external 
particles computed recently \cite{
Berger:2008sj, Giele:2008bc, Berger:2008sz, Ellis:2008qc, 
Lazopoulos:2008ex, Winter:2009kd, vanHameren:2009dr},  provides
a proof of principle that  
reliable description of many $2
\to 4$ processes is  now within reach.

A major reason for extending leading order results to next-to-leading
order is a significant reduction of the unphysical dependence on
factorization and renormalization scales in NLO QCD cross sections and
distributions.  Such reduction is very important, especially for high
multiplicity processes, where the unphysical dependence on scales can
be significant.  Indeed, for such processes, the scale dependence is
amplified by the high power of the strong coupling constant. For a
cross-section involving n jets $\sigma_n \sim \alpha_s^n(\mu)$, so
that small changes in the renormalization scale $\mu$ lead to large
changes in the corresponding cross sections
\be
\mu \frac{\partial \sigma_n }{\partial \mu } \sim
- 2 n \beta_0 \alpha_s \sigma_n,\;\; \beta_0=(11 N_c-2 n_f)/(12 \pi).
\label{eq1}
\ee

There are many cases where NLO computations reduce the dependence of
cross sections on unphysical scales to $10-20\%$ which, given typical
cross section uncertainties at leading order (LO) of about $50\%$, is
a great success. However, even if such scale-independence is observed
for the total-cross section, it is not always possible to claim that a
particular process is described reliably, for the purpose of LHC
phenomenology. The reason is the dual role that many complicated
processes at the LHC will play.  Indeed, depending on the cuts on the
final states, processes that involve e.g. top quarks and/or
electroweak gauge bosons and QCD jets are treated as either primary
signals or unwanted backgrounds and very different cuts are applied to
final states in the two cases.  We will refer generically to these
cuts as ``signal'' or ``background'' cuts,
respectively~\footnote{Since we are mainly concerned with the process
$W+3$ jets, we call ``signal cuts'' the ones where this process is
considered a signal, and ``background cuts'' the ones for which this
process is an unwanted background to some other New Physics signal. We
caution the reader that these terms are often used in the literature
in exactly the opposite way.}
These cuts force final state particles to live in different regions of
phase-space, so that {\it a'priori} it is not possible to relate QCD
corrections to the same process subject to either ``signal'' or
``background'' cuts.

\begin{figure}[t]
\begin{center}
\includegraphics[angle=0,scale=0.6]{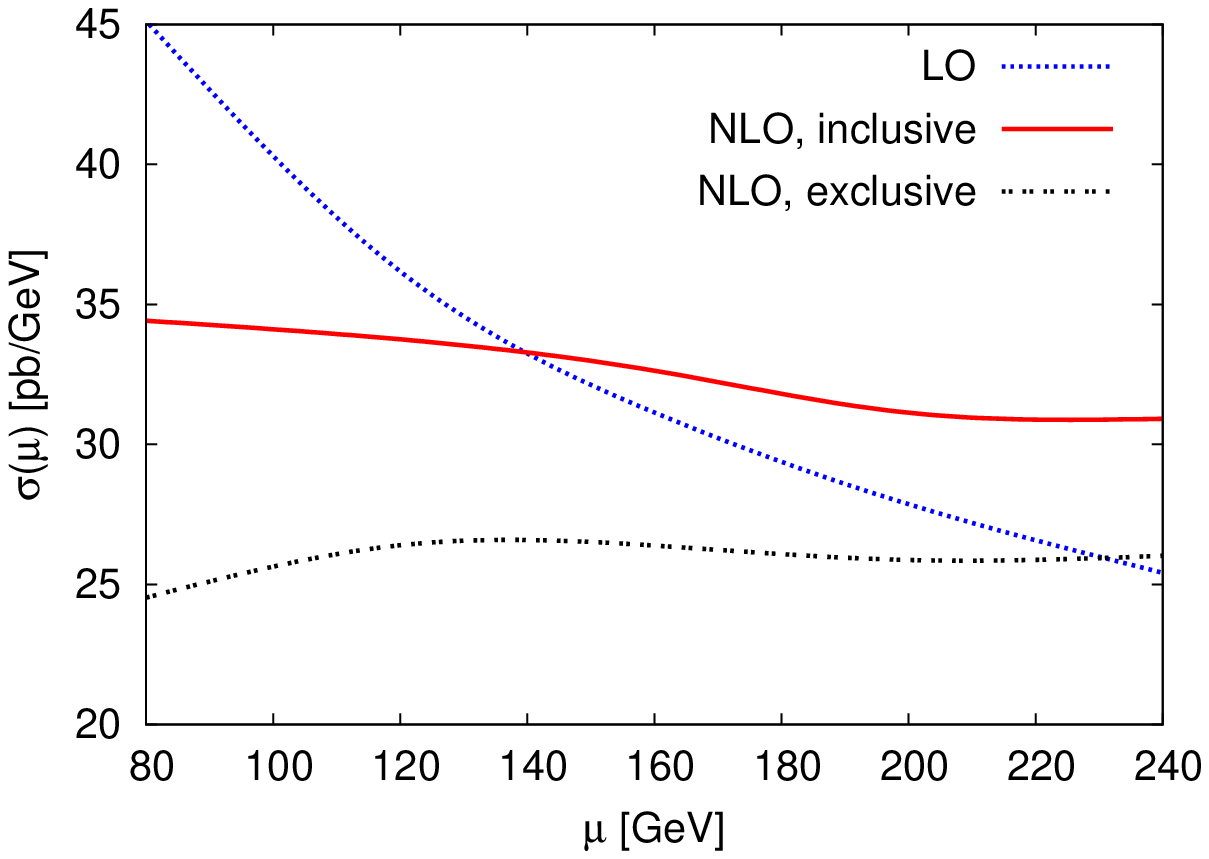}
\caption{
The dependence of the $W^++3~{\rm jet}$ inclusive
production cross section at the LHC on the factorization and  
renormalization
scale $\mu$.  All cuts and parameters are
described in the text. The leading color adjustment procedure is
applied.}
\label{fig:scale}
\end{center}
\end{figure}

A candidate procedure for dealing with LHC processes which have this
dual role is as follows. One starts the study of these processes by
applying signal cuts and using the resultant data set to refine the
theoretical tools, e.g. Monte Carlo event generators.  Once a good
understanding of a given process is achieved with signal cuts, one
uses the refined tools to extrapolate to a different kinematic
situation, specified by the background cuts. Unfortunately, the
reliability of this extrapolation is not assured.  The purpose of
applying the background cuts is to suppress, as far as possible, the
very kinematic configurations that are allowed by signal cuts.
Therefore, such an extrapolation can only work if the influence of
kinematics on QCD radiative effects is correctly captured by the
available tools.  Since only relatively simple theoretical tools, such
as leading order parton integrators or parton showers, are currently
available for complicated final states, the modeling of the radiative
effects is only approximate.  On the other hand, if a NLO QCD
computation is available, such an extrapolation can be done with a
smaller ambiguity since all the relevant scales are generated
dynamically in NLO computations, largely independent of the choices
made initially.  For cases with complicated kinematics, this is
clearly indispensable.

In this paper we discuss and illustrate this issue, taking the
production of $W$ bosons in association with three jets at the LHC as
an example.  For definiteness, we choose to consider proton-proton
collisions at $\sqrt{s}=10$~TeV \cite{lhcschedule}.  However, we do
not aim to describe $W+3$ jet production at the LHC in all possible
detail, since knowledge of the exact experimental setup would be
required.  Instead, we look for and try to understand differences
between NLO and LO QCD results for basic observables, for the case of
signal cuts. We point out that it is not always clear which leading
order predictions should be used in those comparisons since different
choices of renormalization and factorization scales affect the leading
order predictions strongly. We therefore compare our results to a
variety of leading order predictions including most advanced ones,
where matrix element computations are matched to parton showers.

We note that NLO QCD corrections to $W$+3~jet production at the LHC
have been studied in great detail recently in
Ref.~\cite{Berger:2009ep}, mostly for signal cuts.  We have checked a
number of results for $W^\pm$ production cross-sections at the LHC,
reported in that reference, and found agreement within a few percent
in all cases considered.  These small differences are 
compatible with the fact that our calculation employs the leading
color approximation with color adjustment procedure, explained in
detail in Ref.~\cite{Ellis:2009bu}, whereas computation in
Ref.~\cite{Berger:2009ep} accounts for complete color dependence.

We also discuss QCD corrections to {\it background cuts}, studied by
the ATLAS \cite{Mangano:2008ha,Yamazaki:2008nm,Yamamoto:2007it} and
CMS collaborations \cite{cmstdr,Spiropulu:1900zz} for SUSY searches at
the LHC.  Such analyses often assume that Standard Model backgrounds
can be measured in SUSY-free regions to fix normalizations and then
employ LO computations to extrapolate to kinematic regions where
supersymmetric signal is expected.  Hence, an implicit assumption in
those analyses is that LO distributions have correct shapes and that
higher-order QCD effects provide a kinematic-independent
renormalization.  We are now in position to check these assumptions
with the explicit NLO QCD computation of $W+3~{\rm jet}$ process for
typical ATLAS and CMS cuts.

The remainder of the paper is organized as follows.  In
Section~\ref{w3jprocess}, we discuss $W+3$ jet production for signal
cuts at the LHC. In Section~\ref{susy} we study $W+3$ jet production
as a background to SUSY searches for two typical sets of cuts close to
those suggested by the ATLAS and CMS collaborations. In
Section~\ref{conc} we present our conclusions.

\begin{figure}[t]
\begin{center}
\includegraphics[angle=0,scale=0.6]{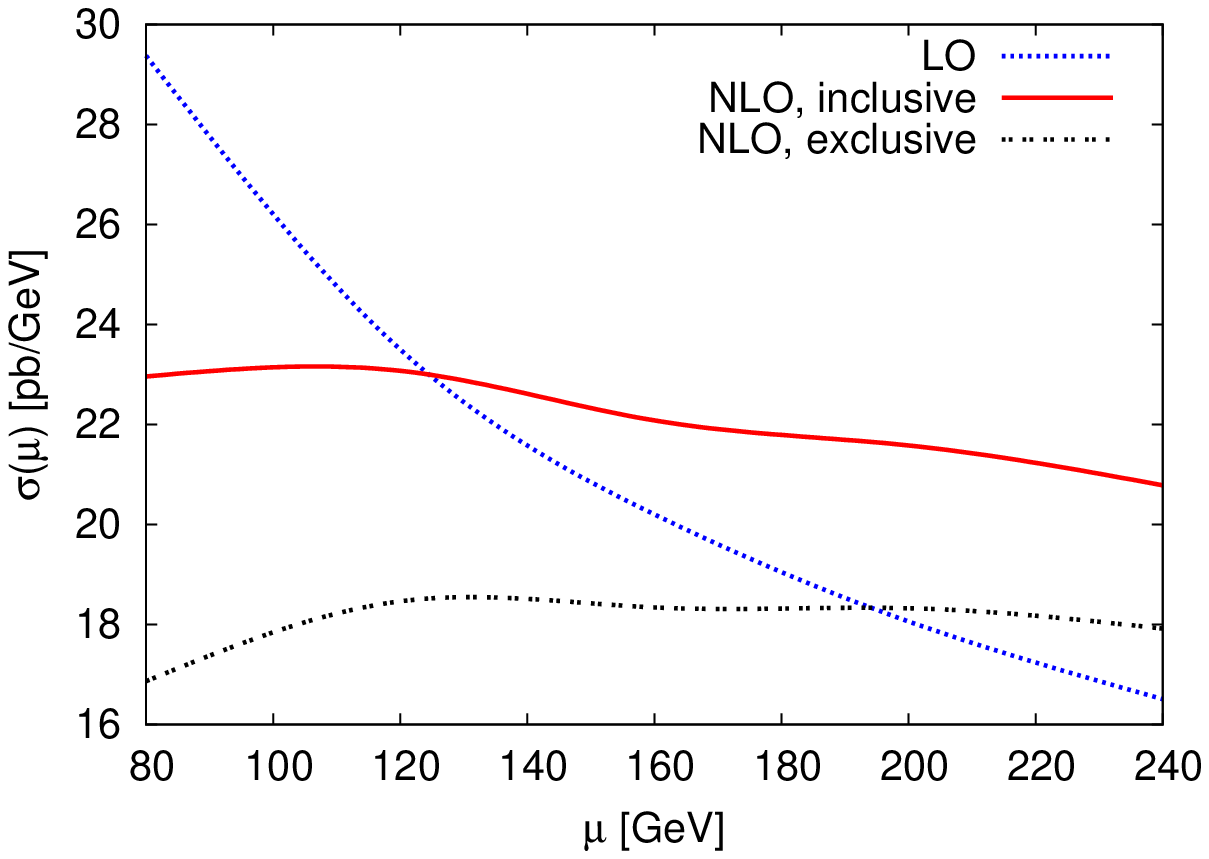}
\caption{
The dependence of the $W^-+3~{\rm jet}$ inclusive
production cross section at the LHC on the factorization and  
renormalization
scale $\mu$.  All cuts and parameters are
described in the text. The leading color adjustment procedure is
applied.}
\label{fig:scaleWm}
\end{center}
\end{figure}

\section{Study of $W+3$ jet process}
\label{w3jprocess}

\begin{figure}[t]
\begin{center}
\includegraphics[angle=0,scale=0.6]{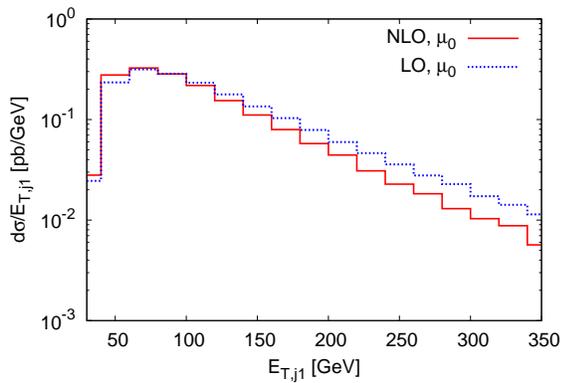}
\caption{The transverse momentum
distribution of the leading jet for $W^++3~{\rm jet}$ inclusive
production cross section at the LHC.  All cuts and parameters are
described in the text. The leading color adjustment procedure is
applied.}
\label{fig:j1}
\end{center}
\end{figure}

In this Section, we discuss NLO QCD effects in $W+3$ jet production
for a set of cuts, designed to study the $W$ production in association
with jets. We follow Ref.~\cite{Ellis:2009bu} closely and perform
calculations in the leading color approximation.  The calculation
relies heavily on the framework provided by MCFM \cite{mcfm} and uses
one-loop amplitudes computed in \cite{Ellis:2008qc}.  We employ the
Catani-Seymour dipole subtraction \cite{cs} to compute real emission
correction; details of our implementation are given in
\cite{Ellis:2009zw}. We use the leading color adjustment procedure
described in that paper to correct for deficiencies of the leading
color approximation, to the extent possible. We note that production
cross-sections for $W^+$ and $W^-$ at the LHC are not the same; we
have chosen to discuss the case of $W^+$ production almost everywhere
in this paper. We do, however, show results for the $W^- + 3$ jet
production cross-section at the LHC in dependence of factorization and
renormalization scales.

We begin by summarizing all the relevant cuts and input parameters
that are employed in the computation.  We take the LHC center-of-mass
energy to be 10 TeV. We require that the transverse momentum and
pseudorapidity of the three jets satisfy $p_{T,j} > 30~{\rm GeV}$ and
$|\eta_j| < 3$. We consider the leptonic decay of the W to electron
(or muon) and employ the following restrictions on lepton transverse
momentum, missing transverse energy, lepton rapidity and $W$-boson
transverse mass, $p_{T,e} > 20~{\rm GeV}$, $\slsh{E}_{T} > 15~{\rm
GeV}$, $|\eta_e| < 2.4$, $M_{T}^W > 30~{\rm GeV}$.  We do not apply an
isolation cut on the leptons.
To define jets, we use the SISCone jet-algorithm~\cite{siscone} with
$R=\sqrt{\Delta \eta^2 + \Delta \phi^2}= 0.5$ and merging parameter $f
= 0.5$.

We consider the production of {\it on-shell} $W^{+}$ bosons, that
decay into a pair of massless leptons. Finite width effects are about
$1\%$; they tend to decrease the cross section.  The CKM matrix is set
equal to the identity matrix; this {\it reduces} the $W+3$ jet
production cross section at the LHC by less than $1\%$. All quarks,
with the exception of the top quark, are considered massless. The top
quark is considered infinitely heavy and its contribution is
neglected. The mass of the $W$ boson is taken to be $m_W = 80.419~{\rm
GeV}$; its couplings to fermions are obtained from $\alpha_{\rm
QED}(m_Z) = 1/128.802$ and $\sin^2
\theta_W = 0.230$.  We use CTEQ6L parton distribution functions for
leading order and CTEQ6M for next-to-leading order computations
\cite{Pumplin:2002vw,Nadolsky:2008zw}.  Note that we do not include
the factor ${\rm Br}(W \to l \nu_l)$ in the results for cross-sections 
quoted below.

We first discuss results for total cross sections. We set
renormalization and factorization scales $\mu$ to $\mu =
[80,120,160,200,240]~{\rm GeV}$ and calculate the cross-sections with
the cuts defined at the beginning of this Section.  The result of the
calculation is illustrated in Fig.\ref{fig:scale}.  For full-color
leading order cross section we find
\begin{equation}
\sigma_{W^++\ge 3j}^{\rm LO,FC} = 35(10)\;{\rm pb}\;,
\label{eq1_1}
\end{equation}
where the $\pm 10~{\rm pb}$ uncertainty from scale variation is shown in 
brackets. Calculating the same cross-section in the 
leading color approximation, 
we find the leading color adjustment parameter 
\begin{equation} 
{\cal R} = 
\sigma_{W^++\ge 3j}^{\rm LO,FC}/\sigma_{W^++\ge 3j}^{\rm LO,LC}
=0.940(5)\,, 
\label{eq:ca}
\end{equation}
where the uncertainty indicates changes in this ratio that we observe
when we change factorization/renormalization scales chosen in leading
order computations or cuts on the final state particles. We also find
that the ${\cal R}$ ratio for the $W^-$ production is the same as for
the $W^+$.  Since ${\cal R}$ does not depend in any significant way on
the details of the process, applied cuts and chosen scales, we use the
central value for ${\cal R}$ given in Eq.(\ref{eq:ca}) in what
follows.

\begin{figure}[t]
\begin{center}
\includegraphics[angle=0,scale=0.6]{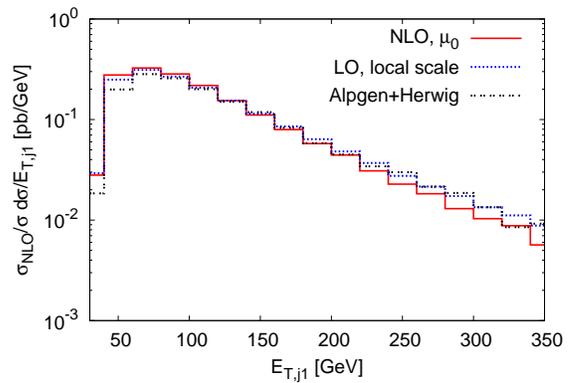}
\caption{The transverse momentum
distribution of the leading jet for $W^++3~{\rm jet}$ inclusive
production cross section at the LHC.  All cuts and parameters are
described in the text. The leading color adjustment procedure is
applied. All LO distributions are rescaled by constant factor, to
ensure that the LO and NLO normalizations coincide.}
\label{fig:j1a}
\end{center}
\end{figure}

At NLO we obtain the {\it adjusted leading-color} 
inclusive cross-section,
$\sigma_{W^++\ge 3j}^{\rm NLO, aLC}({\rm
incl}) = {\cal R} \cdot \sigma_{W^++\ge 3j}^{\rm NLO, LC}({\rm
incl})$,  
\begin{equation}
\sigma_{W^++\ge 3j}^{\rm NLO, aLC}({\rm incl}) = 32.4(1.5)\; 
{\rm pb}\,.
\label{nloxs}
\end{equation}
This result implies (see Fig.\ref{fig:scale}) that for our choice of
cuts and input parameters, NLO QCD corrections to the inclusive
cross-section are very moderate for $\mu \sim 140 - 160~{\rm GeV}$.
We also observe a remarkable reduction in scale dependence from more
than $\pm 30\%$ at leading order to only $\pm 5\%$ at NLO.  While
corrections to the {\it exclusive cross-section } are larger for
similar values of $\mu$, the scale independence of the exclusive NLO
cross-section is similar to the inclusive one.
In Fig.\ref{fig:scaleWm} the cross-section for $W^-+3~$jet production
is shown in dependence on the factorization and renormalization
scales.  The cross section is smaller in this case, while the
stabilization of scale dependence that occurs at next-to-leading order
is very similar for $W^-$ and $W^+$ production cross-sections.

\begin{figure}[t]
\begin{center}
\includegraphics[angle=0,scale=0.6]{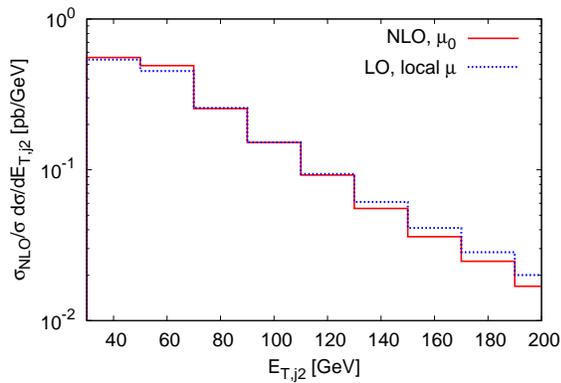}
\caption{The transverse  momentum distribution of the
second hardest jet in $W^++3$ jet production at the LHC. All cuts and
parameters are described in the text. The leading color adjustment
procedure is applied.  The LO distribution is rescaled by constant
factor, to ensure that the LO and NLO normalizations coincide.  }
\label{fig:j2}
\end{center}
\end{figure}

Given that NLO QCD corrections to the total cross sections are small,
it is tempting to surmise that the corrections to kinematic
distributions should also be insignificant.  As we will now show, the
actual situation is more complex.  We consider kinematic distributions
for the inclusive $W^++3$~jet production. We choose to show the NLO
distributions for the {\it dynamical} scale $\mu_0 = \sqrt{p_{T,W}^2 +
m_W^2}$, where $p_{T,W}$ is the transverse momentum of the $W$ boson
as done e.g. in ~\cite{Alwall:2007fs}. We note that for such a scale
the LO cross-section is $\sigma^{\rm LO}_{W^++\ge 3j} = 37.6~{\rm pb}$
and the adjusted leading color NLO cross-section is $\sigma^{\rm NLO,
aLC}_{W^++\ge 3j} = 34.2~{\rm pb}$, consistent with Eq.~(\ref{nloxs})
within the indicated uncertainties.  The radiative corrections to
$W+3~{\rm jet}$ production cross-section at scale $\mu_0$ are
therefore small, about $-10\%$.  For the following discussion, scale
choices in NLO computations are not very important since, as it turns
out, {\it shapes} of NLO distributions are fairly insensitive to them.

 We begin by studying the transverse momentum distribution of the leading 
jet. In Fig.~\ref{fig:j1} we compare NLO and LO predictions for scale
$\mu_0$. We find that the NLO QCD corrections change the shape of this
distribution -- the leading order distribution underestimates the NLO
result at small values of the transverse energy by about $30$ percent
and systematically exceeds the NLO result for higher values of the
transverse energy.  A similar feature is observed in other
distributions related to jet transverse momenta if the NLO result is
compared to LO predictions with the scale $\mu_0$.

\begin{figure}[t]
\begin{center}
\includegraphics[angle=0,scale=0.6]{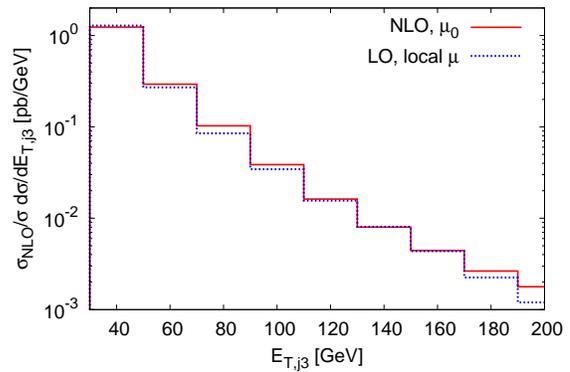}
\caption{The transverse  momentum distribution of the
third hardest jet in $W^++3$ jet production at the LHC. All cuts and
parameters are described in the text. The leading color adjustment
procedure is applied.  The LO distribution is rescaled by constant
factor, to ensure that the LO and NLO normalizations coincide.  }
\label{fig:j3}
\end{center}
\end{figure}

The origin of these shape changes was recently discussed in
Ref.~\cite{Bauer:2009km} using soft-collinear effective theory and in
Refs.~\cite{Ellis:2009bu,Berger:2009zg} in connection with NLO QCD
computations for $W+3~{\rm jets}$ production at the Tevatron and the
LHC. Here, we recapitulate the explanation of the inadequacy of the
scale $\mu_0$ given in Ref.~\cite{Ellis:2009bu}.  This inadequacy is
related to two facts:
a) in the region where the jets have large transverse momentum,
  the $W$-boson transverse momentum spectrum is softer than that of  
the jets;
b) the probability of parton branching is determined by the relative
transverse momentum of the two daughter partons produced in that
branching; such transverse momentum should be the appropriate scale
for the strong coupling constant.  When these two facts are combined,
one is led to the conclusion that in the kinematic region where the
jets have large transverse momenta, the use of $\alpha_s(\mu_0)$ in LO
computations overestimates the cross section.  At next-to-leading
order, the appropriate scale for the strong coupling constant $\mu
\sim p_{T,j} \gg \mu_0$ is generated dynamically and the cross section
in that region becomes smaller.

Is it possible to account for the shape modifications by more
sophisticated LO computations?
 The affirmative answer to this question was
given in Refs.~\cite{Bauer:2009km,Berger:2009zg}, where particular
choices of scales set by e.g. the hadronic invariant mass or total
transverse energy in an event, were advocated. It should be
emphasized, however, that the idea to employ scales of the strong
coupling that are determined from local kinematics on an
event-by-event basis is not new since it is central to both parton
showers and advanced leading order computations that employ matrix
elements and parton shower matching \cite{Catani:2001cc}.

\begin{figure}[t]
\begin{center}
\includegraphics[angle=0,scale=0.6]{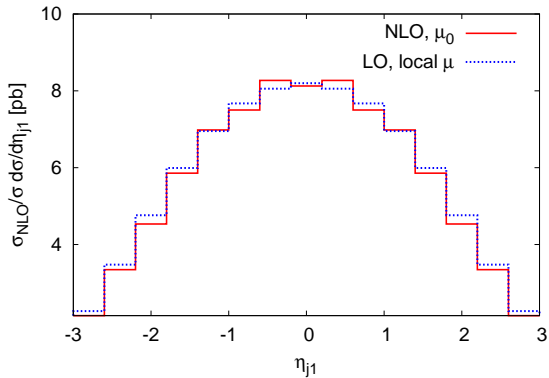}
\caption{The rapidity
distribution of the leading jet for $W^++3~{\rm jet}$ inclusive
production cross section at the LHC.  All cuts and parameters are
described in the text. The leading color adjustment procedure is
applied.  The LO distribution is rescaled by constant factor, to
ensure that the LO and NLO normalizations coincide.  }
\label{fig:rap1}
\end{center}
\end{figure}

Since such ``local'' scales capture the kinematics of complicated
events correctly, it is conceivable that they produce shapes that are
close to exact NLO results.  We show the comparison of the NLO
prediction with two leading order results in Fig.\ref{fig:j1a}.

One LO distribution is obtained by following the MLM procedure
 whose application to $W+3~{\rm jet}$ production
is described in Ref.~\cite{Alwall:2007fs}. The MLM procedure and its
close relative the CKKW algorithm \cite{Catani:2001cc} are the most
advanced techniques available currently for leading order predictions,
so it is interesting to see how it compares with NLO computations.
We use Alpgen \cite{alpgen} to generate unweighted events that are
matched to the Herwig \cite{herwig} parton shower.  We produce hard
events with up to five QCD partons in the final state with Alpgen,
using a transverse momentum cut of $p_{\rm tj,min}= 20~(25)$ GeV and a
separation parameter $d_{\rm rj}= 0.35~(0.45)$
\cite{alpgen}. To shower the hard events with Herwig we used 
$R_{\rm clus} = d_{rj}$ and $E_{\rm t,clus} = p_{\rm tj, min}$ as
matching parameters for the MLM prescription \cite{alpgen}.  We find
that results are fairly independent of the cuts used in the generation
of the hard events and that samples with five hard partons contribute
little. This indicates that hard samples with yet higher multiplicity
can be safely neglected.

\begin{figure}[t]
\begin{center}
\includegraphics[angle=0,scale=0.6]{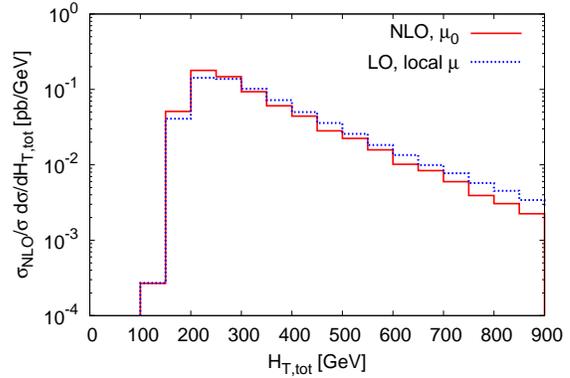}
\caption{The total transverse energy
distribution for $W^++3~{\rm jet}$ inclusive production cross section
at the LHC.  All cuts and parameters are described in the text. The
leading color adjustment procedure is applied.  The LO distribution is
rescaled by constant factor, to ensure that the LO and NLO
normalizations coincide.  }
\label{fig:ht}
\end{center}
\end{figure}

The other LO prediction shown in Fig.~\ref{fig:j1a} is our
implementation of the local scales in the strong coupling constant; it
is close in spirit to the re-weighting part of the CKKW procedure
\cite{Catani:2001cc}.  To this end, for a given LO partonic event that
passes jet cuts, we cluster partons according to the measure given by
$k_\perp$-jet algorithm\footnote{We note that the jet cuts can be
defined with any jet algorithm; the $k_\perp$-algorithm is only used
to reconstruct the event branching history.}.  A repeated clustering
gives us a ``branching history'' that can be associated with the
event; at each branching the scale of the strong coupling constant is
 chosen as the relative momentum of two daughters in the branching.
We will refer to scales of the strong coupling constant chosen by this
algorithm as ``local'' scales. Note that this procedure is strictly a
simple way to set scales of the strong coupling constant to reasonable
values in $W+5~{\rm parton}$ leading order matrix elements. In doing
so, we do not try to combine matrix elements of different
multiplicities nor do we attempt to shower leading order partonic
configuration. Differences between distributions produced with Alpgen
and with the local scale procedure give an idea of the importance of
the parton shower and Sudakov re-weighting.

We point out that such modifications of leading order computations may
lead to large changes in the cross-sections. For example, Alpgen
cross-section is $\sim 22~{\rm pb}$ and the local scale cross-section
is $\sim 47~{\rm pb}$, to be compared with $\sim 33~{\rm pb}$ NLO
cross-section.  However, the normalization of cross-sections is a hard
problem where next-to-leading computations or direct normalization to
data are the only known solutions. To separate issues of normalization
from the shape, we normalize all leading order results in
Fig.~\ref{fig:j1a} to the NLO cross-section.  We observe that both the
Alpgen+Herwig distribution and the local scale distribution describe
the NLO result fairly well. Also, the proximity between the shapes of
the two leading order results tells us that parton shower does
relatively little to alter the shape of the distribution.

\begin{figure}[t]
\begin{center}
\includegraphics[angle=0,scale=0.6]{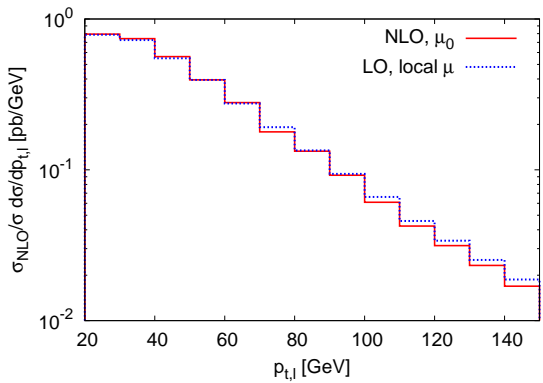}
\caption{The transverse momentum distribution of the charged lepton
for $W^++3~{\rm jet}$ inclusive production cross section at the LHC.
All cuts and parameters are described in the text. The leading color
adjustment procedure is applied. The LO distribution is rescaled by
constant factor, to ensure that the LO and NLO normalizations
coincide.  }
\label{fig:lept1}
\end{center}
\end{figure}

\begin{figure}[t]
\begin{center}
\includegraphics[angle=0,scale=0.6]{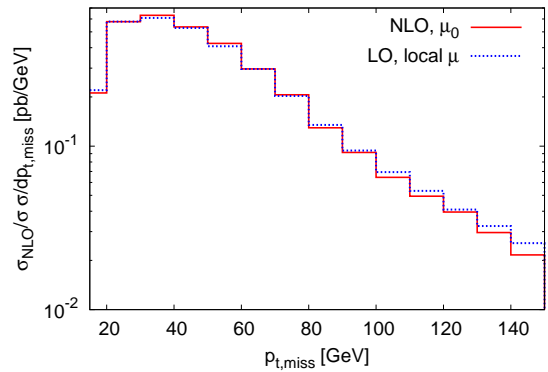}
\caption{
The missing transverse momentum distribution for $W^++3~{\rm jet}$
inclusive production cross section at the LHC.  All cuts and
parameters are described in the text. The leading color adjustment
procedure is applied.  The LO distribution is rescaled by constant
factor, to ensure that the LO and NLO normalizations coincide.  }
\label{fig:lept2}
\end{center}
\end{figure}

We find that these observations are generic: leading order
computations obtained with either Alpgen+Herwig or local scales are
similar and they work reasonably well in reproducing shapes of NLO 
distributions. We believe this is important conclusion, especially 
in the case of Alpgen+Herwig since those programs are used by 
experimenters as tools for understanding properties of $W+$ jets process.
In order to avoid too busy plots, 
we choose to only  show leading order results computed with the ``local''
scale choice for the strong coupling constant in what follows. 
We stress that for all distributions the normalization of the leading
order cross-section is adjusted to agree with next-to-leading order
result.  We show the distribution in transverse energy of the
second-hardest and third-hardest jet in
Figs.~\ref{fig:j2},\ref{fig:j3}, rapidity of the hardest jet in
Fig.~\ref{fig:rap1} and the distribution in total transverse energy
$
H_{\rm T,tot} = \sum_{\rm jets} |p_{\rm T,j}| + p_{\rm T,l} + \slsh{p}_T
$
in Fig.\ref{fig:ht}.  We also show leptonic distributions in
Figs.~\ref{fig:lept1} and \ref{fig:lept2} where we plot the lepton
transverse momentum and the missing transverse momentum, respectively.
As stated, in all considered cases local scales reproduce shapes of
the distributions quite well.

\section{$W+3$ jet production as a model for background to
supersymmetric searches}
\label{susy}

In this Section we investigate QCD corrections to $W+3$ jet production
at the LHC for a set of cuts appropriate in supersymmetric
searches. By construction, these background cuts seek to suppress the
production of $W$ bosons in association with jets as much as possible,
effectively driving $W+{\rm jet}$ production to corners of the
available phase-space. It is therefore unclear if QCD radiative
effects in those regions of phase-space are similar to QCD corrections
to the production cross-sections discussed in the previous Section.
To answer this question, we discuss two types of cuts, very similar to
those suggested by the ATLAS and CMS collaborations, in their planned
searches for supersymmetry at the LHC.~\footnote{We point out that we
kept cuts very similar to those used in the experimental studies done
at $14$TeV despite the fact that we use $10$ TeV as center-of-mass
energy. As a consequence cross-sections in this section are very
small. A more realistic study would require adapting those cuts to the
centre-of mass energy, but this is beyond the scope of this paper.}
All the input parameters are the same as in the previous Section
except that we use merging parameter $f = 0.7$ to define jets using
SIScone algorithm.

\subsection{ATLAS setup -- faking jets from $\tau$ decays}
\label{sec:cutsII}
We begin by considering cuts employed by the ATLAS collaboration to
search for SUSY with $R$-parity conservation.  In that case, the
typical signal comes from gluino pair-production. If each gluino
decays into two jets and a neutralino, a SUSY signature will involve 4
jets and missing transverse energy. A dominant background to this
process comes from $Z+4$ jet production, with the subsequent decay of
the $Z$-boson into two neutrinos.  Another important background comes
from $W^++3$ jet production\footnote{Clearly, there is also a similar
background from $W^-+3$ jet production but we do not consider it
here.}, followed by the decays $W^+ \to \bar \tau
\nu_\tau \to \bar \nu_\tau \nu_\tau +{\rm hadrons}$, so that hadrons
from semileptonic decay of the $\tau$ lepton produce the fourth jet.

One can use peculiar kinematic properties of the fourth jet to connect
it to $\tau$ decays and then reject such events but, because of
limited efficiency in identifying $\tau$ decays and because the cross
section for $W+3$ jet production is almost two orders of magnitude
larger than the $Z+4$ jet production cross section, it is important to
consider this source of the background as well.

We begin by listing a typical set of cuts that the ATLAS collaboration
applies to suppress the $W \to \tau + 3j$
background~\cite{Yamazaki:2008nm,Yamamoto:2007it,Mangano:2008ha}.
First, all jets are required to have transverse momenta larger than
$50~{\rm GeV}$ and the transverse momentum of the leading jet should
exceed $100$~{\rm GeV}. Second, missing energy in the event should
satisfy $\slsh{E}_{T} > \max(100\;{\rm GeV}, 0.2 H_{T})$ with $H_{T} =
\sum_{j} p_{T,j} +\slsh{E}_{T}$.  Third, no leptons with transverse
momenta higher than $20~{\rm GeV}$ should be present. Fourth, jets
should be central $|\eta_j| < 3$.  Finally, the event is required to
be spherical and the cut $S_{\rm T} > 0.2$ is applied on the
transverse sphericity. We will not employ the sphericity cut in what
follows because this observable is not collinear safe at parton
level. In addition, since we consider semileptonic decays of the
$\tau$ lepton, no high-$p_T$ lepton is present in our events and we do
not need to employ a $20~{\rm GeV}$ lepton cut.  The primary
observable is the distribution in the effective mass $H_{\rm T}$
defined above and the range of a particular interest for SUSY
searches, given existing bounds on gluino masses, is $ H_{\rm T} \gsim
1~{\rm TeV}$.

A clear exposition of the effect that the ATLAS cuts have on $W \to
\tau + 3j$ background at leading order was recently given in
Ref.~\cite{Mangano:2008ha}.  It turns out that these cuts primarily
change the normalization of the background but do not significantly
affect the shape of the effective mass distribution, especially in the
region $H_{T} \gsim 1~{\rm TeV}$.  We would like to understand the
impact of NLO QCD corrections to $W \to \tau+3$ jet on the $H_{T}$
distribution.  Our implementation of radiative corrections
incorporates $W$ decay to any leptonic final state but subsequent
hadronic decays of the $\tau$-lepton are not included.  Yet, as we
will argue now, this is not necessary if all we need is an estimate of
the QCD effects.

We note that, given the above cuts and, in particular, the cut on the
missing transverse energy, the $\tau$ lepton produced in $W$ decays
will be highly boosted and its decay products will be very
collimated. We then completely neglect the angular distribution of the
$\tau$ decay products and assume a perfect collinear splitting.  If,
in addition, we neglect all the spin correlations in $\tau$ decay
$\tau \to \nu_\tau q_i \bar q_j$, we conclude that the neutrino has to
carry away about a third of the $\tau$ momentum while the hadronic jet
formed by a quark and an anti-quark from $\tau$ decay has to carry
away two-thirds of the original $\tau$ momentum.  We also expect that, 
since the $\tau$ lepton  is highly boosted, all its hadronic decay channels 
will contribute to the same jet, making the  inclusive treatment of jet 
properties a reasonable approximation. 

\begin{figure}[t]
\begin{center}
\includegraphics[angle=0,scale=0.6]{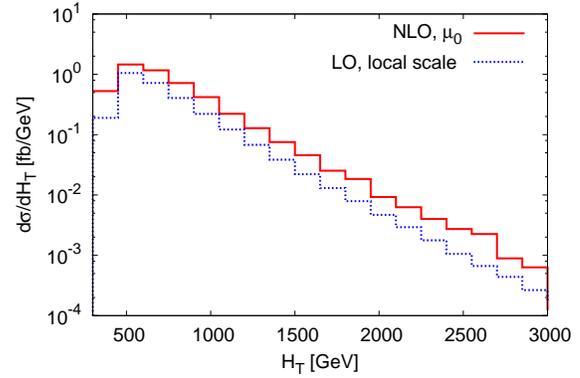}
\caption{Distributions in effective mass $H_{T}$
for $(W^+ \to \bar \tau) + 3 $ jet sample for ATLAS SUSY cuts
described in the text.  All cuts and parameters are described in the
text. The leading color adjustment procedure is applied. The large
difference between LO and NLO distributions can be absorbed by
re-scaling the LO distribution by a constant factor.
}
\label{fig:II1}
\end{center}
\end{figure}

We can implement this set up in our calculation by producing a $W$
boson and letting it decay to a massless lepton and a massless
neutrino.  We then carry through all the steps required for the NLO
QCD computation until the moment when the kinematics of events is
examined and weights, relevant for various histogram bins, are
calculated. At this point, we assign one-third of the lepton momentum
to additional missing energy carried away by $\nu_\tau$ and two-thirds
of the lepton momentum to the fourth ($\tau$) jet in the event.  Note
that we do not apply the jet algorithm to check whether or not the
hadronic jet from $\tau$ decay is sufficiently separated from the
other three jets.~\footnote{We did however impose a separation
$R_{lj}=0.5$ between the $\tau$-lepton and the jets.}  Since this step
is not necessary for infra-red safety, we feel that it is entirely
justified to omit it, given the approximate nature of our
analysis. For next-to-leading computations, we use the leading color
adjustment procedure; we find that ${\cal R} = 0.93$ is an appropriate
value of the re-scaling parameter for ATLAS cuts.

The results of our computation are presented in Fig.~\ref{fig:II1}
where the LO $H_{\perp}$ distribution for our default local scale
 compared to the NLO distribution for the factorization and
renormalization scales set to $\mu_0$.  We point out that the shape of
the leading order distribution is similar to that obtained with Alpgen
presented in Ref.~\cite{Mangano:2008ha}, especially at high values of
the effective mass. At lower values of the effective mass, there is a
dependence on the modeling of $\tau \to {\rm hadrons}$ transition and,
given the very approximate nature of our procedure, it is not
surprising that it tends to fail. It is reassuring, however, that our
procedure seems to work quite well for high values of the effective
mass.

As follows from Fig.~\ref{fig:II1}, the $H_T$ mass distribution
receives {\it large positive} QCD corrections for ATLAS cuts.  Note
that distributions for local scales are {\it not} normalized to match
the NLO distribution there.
We studied the scale dependence of the leading order predictions by
varying local scales around the central value by a factor of two.
While we observe large $\sim \pm 50$\% scale dependence in the LO
result, the NLO QCD corrections are $\sim 100$\% and are thus
considerably larger than what the LO scale variation suggests. The
scale dependence of the $H_\perp$ distribution does decrease
considerably at NLO.
We find that NLO QCD effects provide a universal enhancement of $H_T$
distribution without distorting its shape.  Interestingly, the cuts on
jets and missing energy presented at the beginning of this Section
have a similar impact on the $(W \to \tau) + 3$ jet background -- each
of the individual cuts reduces the magnitude of $(W \to \tau) + 3$ jet
by a factor between three and four, without affecting the shape of the
$H_T$ distribution~\cite{Mangano:2008ha}.  NLO QCD effects therefore
are comparable to the effects of the cuts and work {\it in the
opposite direction}.

We emphasize that, had we chosen scale $\mu_0$ also in LO computation,
we would observe large {\it positive} NLO QCD effects for $H_\perp$
distribution, in sharp contrast with large {\it negative} corrections
for such scale choice in high-$p_{\perp, j}$ regions, described in the
previous Section (see Fig.~\ref{fig:j1}).  This is not surprising
since, in contrast to $W+3$ jet signal cuts, ATLAS cuts require large
amount of missing energy, which forces $W$ transverse momentum to be
comparable or larger than transverse momenta of hard jets in the
event.  Jet branching on the other hand, can occur at lower relative
transverse momenta.  Taking the relative transverse momentum as the
correct scale for the strong coupling constant, it is natural that LO
cross sections for $\mu = \mu_0$ strongly underestimate the $H_T$
distribution. This is indeed what we see when LO and NLO results are
compared.

We believe that this discussion shows explicitly how problematic
extrapolation from signal to background region can be since the NLO
QCD effects for ATLAS cuts have no relation whatsoever to the NLO QCD
effects for the total cross section. This mismatch happens because the
kinematic region selected by ATLAS cuts gives negligible contribution
to the total cross section. On the other hand, it appears that one can
use low $H_T < 1~{\rm TeV}$ bins for ATLAS cuts to fix background
normalization since QCD effects seem to be $H_T$-independent and SUSY
contamination in low-$H_T$ bins is small.

\subsection{CMS indirect lepton veto cut} 

How robust is the situation discussed in connection with ATLAS cuts?
To answer this question, we study another example of background cuts.
Those cuts are adopted by the CMS collaboration for SUSY searches at
the LHC~\cite{cmstdr,Spiropulu:1900zz}.  The target signal is gluino
pair production and the final state involves jets and missing
transverse energy.

The CMS collaboration does not veto leptons directly.  Rather, cuts
are designed in such a way that the contribution of $W+{\rm jets}$
becomes naturally small.  Such cuts are usually referred to as {\it
indirect lepton veto} cuts. We approximate the CMS indirect lepton
veto cut by requiring that there are three or more jets in the
event. The missing energy in the event should be large, $E_{\rm miss}
> 200~{\rm GeV}$.  The leading jet in the event should be very central
$|\eta_{\rm lead~jet}| < 1.7$ while all other jets should be in the
central region $|\eta_{\rm other~jets}| < 3$. Jets are defined with
the transverse momentum cut of $p_{T,j} > 30~{\rm GeV}$ but the
transverse momentum of the leading and sub-leading jets should be
larger that $180$ and $110$~{\rm GeV}, respectively.  Leptons from $W$
decays should satisfy the same cuts as jets but lepton transverse
momentum can not be the largest or next-to-largest in a particular
event; experimentally, this requirement is implemented by cutting on
the fraction of electromagnetic energy carried by a ``jet''.
  Finally, a particular effective mass
is required to be large
$H_{{\rm T},24} =
\sum \limits_{j=2}^{4} p_{T,j}
+E_{\rm miss} > 500~{\rm GeV}$. To calculate the sum in this formula one
orders leptons and jets according to their hardness, disregards the
leading jet and sums over transverse momenta of second-to-leading,  
third-to-leading
and fourth-to-leading particles/jets.

\begin{figure}[t]
\begin{center}
\includegraphics[angle=0,scale=0.6]{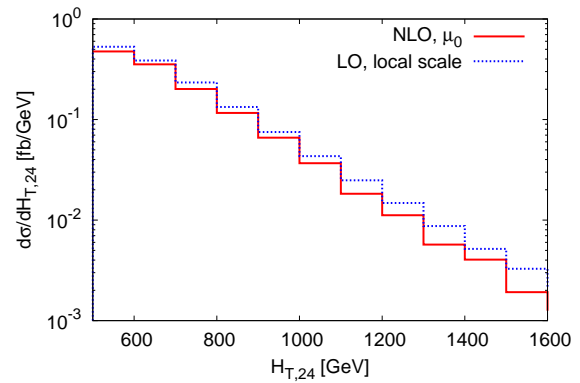}
\caption{Distributions in reduced transverse mass $H_{T, 24}$
for $W^++3$ jet events with CMS SUSY cuts that define indirect lepton
veto procedure as described in the text.  All cuts and parameters are
described in the text. The leading color adjustment procedure is
applied.
}
\label{fig:III1}
\end{center}
\end{figure}

\begin{figure}[t]
\begin{center}
\includegraphics[angle=0,scale=0.6]{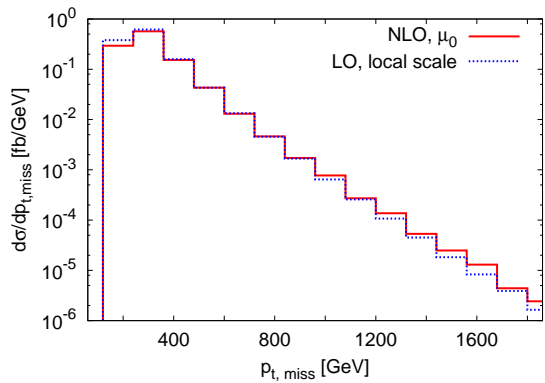}
\caption{Distributions in the missing transverse energy for $W^++3$ jet events with CMS SUSY cuts that define
indirect lepton veto procedure as described in the text.  All cuts and
parameters are described in the text. The leading color adjustment
procedure is applied.
}
\label{fig:III2}
\end{center}
\end{figure}

We show the result of our computation of the NLO QCD corrections to
the $W+3$ jet cross section in case of CMS-style cuts in
Figs.~\ref{fig:III1},\ref{fig:III2}, where distributions in $H_{{\rm
T},24} $ and missing energy are plotted.  We again use $\mu_0$, as the
factorization and renormalization scales and vary it by a factor two
up and down to estimate scale uncertainties. The NLO corrections for
these cuts change the LO result by $-40\%$ to $-10 \%$ depending on
the scale chosen in LO computations. For $\mu = \mu_0$, the
corrections are about $-10\%$ and no significant changes of shape are
observed.  In this case, the scale variation at leading order gives a
good indication of the size of NLO QCD corrections.

It is striking that the magnitude of NLO QCD corrections for CMS cuts
is in strong contrast with the magnitude of NLO QCD effects for ATLAS
cuts, discussed in the previous Section. This emphasizes the
dependence of NLO QCD corrections on exact implementation of kinematic
cuts even if such cuts are designed to target very similar physics
beyond the Standard Model. On the other hand, we find that shapes of
basic distributions employed in supersymmetric searches are described
fairly well by leading order computations, for both ATLAS and CMS
cuts.  If one can verify that, say, low-$H_\perp$ bins are not
contaminated by New Physics, those bins can be used to determine the
normalization of the background.

\section{Conclusions}
\label{conc}

We have discussed the NLO QCD corrections to $W+3$ jet production at
the LHC.  We found that the inclusion of NLO QCD corrections leads to
a significant reduction in dependence of LO results on the
renormalization and factorization scales; the residual uncertainty
associated with the total cross section is $\pm 5\%$.  We showed that
small corrections to total cross sections do not necessarily imply
that corrections to differential distributions are small and there is
a high degree of non-uniformity in these corrections across the
available phase-space.

It should be stressed that the last statement depends upon
renormalization and factorization scales chosen in leading order
computations.
In particular if leading order calculations are done with the scale
$\mu =\sqrt{p_{T,W}^2+m_W^2}$ we find a large difference in shapes
between LO and NLO distributions.  On the other hand, it is clear {\it
a'priori} that better results are achievable if scales are chosen
based on local probabilities for jet branching.
Here we have shown explicitly that when a local scale choice for the
strong coupling constant is employed in leading order computations,
such computations reproduce shapes of various NLO distributions quite
well.  Note that any leading order computation matched to parton
shower in the spirit of CKKW procedure \cite{Catani:2001cc} {\it does}
employ such local scales and our NLO analysis therefore confirms that,
as far as shapes of various kinematic distributions are concerned,
this is a very reasonable procedure.

The production of $W$-bosons in association with three jets is an
important background for SUSY searches in jets + missing energy
channels.  We studied NLO QCD corrections to cuts employed by ATLAS
and CMS collaborations for SUSY searches and found that such
corrections are not at all correlated with corrections to the total
cross sections.  It is peculiar that the magnitude of NLO QCD
corrections to, say, effective transverse mass distributions, is very
different for ATLAS and CMS cuts in spite of the fact that these cuts
are designed to serve the same purpose. We find large ($\sim 100 \%$)
corrections for ATLAS and small $(\sim 10\%)$ QCD corrections for CMS
cuts.  We believe that this non-uniformity of corrections and their
apparent strong dependence of the experimental set-up emphasizes the
need for extending NLO QCD studies to other relevant backgrounds such
as $W+4$ jets and $Z+3,4$ jets.  We hope that techniques for NLO QCD
computations developed in recent years make such computations
possible.

{\bf Acknowledgments}
We would like to thank Keith Ellis for collaboration at the early  stages 
of this work,  useful discussions and advice. 
We are grateful  to 
Zoltan Kunszt, Fabio Maltoni, Michelangelo Mangano, Fulvio Piccinini and Gavin Salam
for useful discussions and correspondence.
Parts of this work were carried our during stays at 
Fermilab and CERN whose support and hospitality we gratefully 
acknowledge.
K.M. is supported by NSF under grant PHY-0855365 and by the start up
funds provided by Johns Hopkins University. G.Z. is supported by the
British Science and Technology Facilities Council.
Calculations reported in this paper were performed on the Homewood
High Performance Cluster of Johns Hopkins University.

\end{document}